\documentclass[12pt,a4j]{article} 
\makeatletter
\newdimen{\eqarcolsep}
\usepackage[dvips]{color,graphicx} 
\usepackage{amsmath} 
\usepackage{amsthm} 
\usepackage{bm} 
\usepackage{fancybox} 
\usepackage{ascmac} 
 \usepackage{amssymb}

\makeatletter 
\begin{document}

{\large \bf General theory of regular biorthogonal pairs and its physical applications}

\begin{center}
Hiroshi Inoue \\
\end{center}

\begin{abstract} 
In this paper we introduce a general theory of regular biorthogonal sequences and its physical applications. Biorthogonal sequences $\{ \phi_{n} \}$ and $\{ \psi_{n} \}$ in a Hilbert space ${ \cal H}$ are said to be regular if $Span\; \{ \phi_{n} \}$ and $Span\; \{ \psi_{n} \}$ are dense in ${\cal H}$. The first purpose is to show that there exists a non-singular positive self-adjoint operator $T_{\bm{f}}$ in ${\cal H}$ defined by an ONB $\bm{f} \equiv \{ f_{n} \}$ in ${\cal H}$ such that $\phi_{n}=T_{\bm{f}} f_{n}$ and $\psi_{n}= T_{\bm{f}}^{-1} f_{n}$, $n=0,1, \cdots$, and such an ONB $\bm{f}$ is unique. The second purpose is to define and study the lowering operators $A_{\bm{f}}$ and $B_{\bm{f}}^{\dagger}$, the raising operators $B_{\bm{f}}$ and $A_{\bm{f}}^{\dagger}$, the number operators $N_{\bm{f}}$ and $N_{\bm{f}}^{\dagger}$ determined by the non-singular positive self-adjoint operator $T_{\bm{f}}$. These operators connect with ${\it quasi}$-${\it hermitian \; quantum \; mechanics}$ and its relatives. This paper clarifies and simplifies the mathematical structure of this framework minimized the required assumptions.\\
\end{abstract}

\section{Introduction}
In this paper we introduce a general theory of regular biorthogonal sequences and its physical applications. Sequences $\{ \phi_{n} \}$ and $\{ \psi_{n} \}$ in a Hilbert space ${\cal H}$ are biorthogonal if $( \phi_{n} | \psi_{m})= \delta_{nm},$ $n=0,1, \cdots,$ where $( \cdot | \cdot )$ is an inner product of ${\cal H}$ and they are regular if both $Span\; \{ \phi_{n} \}$ and $Span \; \{ \psi_{n} \}$ are dense in ${\cal H}$. Then $( \{ \phi_{n} \} , \{ \psi_{n} \} )$ is said to be a regular biorthogonal pair.
The first purpose is to show that the following statements (i)-(iii) are equivalent:\\
\hspace{3mm} (i) $( \{ \phi_{n} \} , \{ \psi_{n} \} )$ is a regular biorthogonal pair in a Hilbert space ${\cal H}$.\\
\hspace{3mm} (ii) For any ONB $\bm{e} = \{e_{n} \}$ in ${\cal H}$, there exists a densely defined closed operator $T$ in ${\cal H}$ with densely defined inverse such that $ \{ e_{n} \} \subset D(T) \cap D((T^{-1})^{\ast})$, $\phi_{n}=Te_{n}$ and $\psi_{n} =(T^{-1})^{\ast} e_{n}$, $n=0,1, \cdots$ and the minimum in such operators $T$ exists and denoted by $T_{\bm{e}}$.\\
\hspace{3mm} (iii) There exists a unique ONB $\bm{f}= \{ f_{n} \}$ in ${\cal H}$ such that $T_{\bm{f}}$ is a non-singular positive self-adjoint operator in ${\cal H}$.\\\\
 Furthermore, we investigate the relationship between a regular biorthogonal pair $( \{ \phi_{n} \} , \{ \psi_{n} \} )$ and the notions of Riesz bases and semi-Riesz bases. Here $( \{ \phi_{n} \} , \{ \psi_{n} \} )$ is a pair of Riesz bases if there exists an ONB $\bm{e}= \{ e_{n} \}$ in ${\cal H}$ such that both $T_{\bm{e}}$ and $T_{\bm{e}}^{-1}$ are bounded. And $( \{ \phi_{n} \} , \{ \psi_{n} \} )$ is a pair of semi-Riesz bases if there exists an ONB $\bm{e}= \{ e_{n} \}$ in ${\cal H}$ such that either $T_{\bm{e}}$ or $T_{\bm{e}}^{-1}$ are bounded. It is shown that Riesz bases and semi-Riesz bases do not depend on methods of taking ONB. In Ref. \cite{h-t}, we have defined and studied the notion of generalized Riesz bases:
Biorthogonal sequences $\{ \phi_{n} \}$ and $\{ \psi_{n} \}$ in a Hilbert space ${\cal H}$ is said to be a generalized Riesz base if there exist an ONB $\bm{e}= \{ e_{n} \}$ in ${\cal H}$ and a densely defined closed operator $T$ in ${\cal H}$ with densely defined inverse such that $\bm{e} \subset D(T) \cap D((T^{-1})^{\ast})$, $\phi_{n}=T e_{n}$ and $\psi_{n}=(T^{-1})^{\ast} e_{n},$ $n=0,1, \cdots $, and then$\{ \phi_{n} \}$ is said to be a Riesz base if the above $T$ and $T^{-1}$ are bounded. As known in the above (ii), in this paper we show that if $( \{ \phi_{n} \} , \{ \psi_{n} \} )$ is a regular biorthogonal pair, then it is always a generalized Riesz base.
The second purpose is to define and study the physical operators determined by the non-singular positive self-adjoint operator $T_{\bm{f}}$ as follows:
\begin{eqnarray}
A_{\bm{f}}
&=& T_{\bm{f}} \left( \sum_{k=0}^{\infty} \sqrt{k+1} f_{k} \otimes \bar{f}_{k+1} \right) T_{\bm{f}}^{-1} , \nonumber \\
B_{\bm{f}}
&=& T_{\bm{f}} \left( \sum_{k=0}^{\infty} \sqrt{k+1} f_{k+1} \otimes \bar{f}_{k} \right) T_{\bm{f}}^{-1}, \nonumber 
\end{eqnarray}
\begin{eqnarray}
A_{\bm{f}}^{\dagger}
&=& T_{\bm{f}}^{-1} \left( \sum_{k=0}^{\infty} \sqrt{k+1} f_{k+1} \otimes \bar{f}_{k} \right) T_{\bm{f}}, \nonumber 
\end{eqnarray}
\begin{eqnarray}
B_{\bm{f}}^{\dagger}
&=& T_{\bm{f}}^{-1} \left( \sum_{k=0}^{\infty} \sqrt{k+1} f_{k} \otimes \bar{f}_{k+1} \right) T_{\bm{f}},  \nonumber \\
N_{\bm{f}}
&=& T_{\bm{f}} \left( \sum_{k=0}^{\infty} \sqrt{k+1} f_{k+1} \otimes \bar{f}_{k+1} \right) T_{\bm{f}}^{-1}, \nonumber \\
N_{\bm{f}}^{\dagger}
&=& T_{\bm{f}}^{-1} \left( \sum_{k=0}^{\infty} \sqrt{k+1} f_{k+1} \otimes \bar{f}_{k+1} \right) T_{\bm{f}}, \nonumber 
\end{eqnarray}
where the tensor $x \otimes \bar{y}$ of elements $x, \; y$ of ${\cal H}$ is defined by
\begin{eqnarray}
(x \otimes \bar{y})\xi =( \xi | y)x, \;\;\; \xi \in {\cal H}. \nonumber
\end{eqnarray}
Then it is shown that these operators defined by ONB do not depend on methods of taking ONB and
\begin{eqnarray}
A_{\bm{f}} \phi_{n}
&=& \left\{
\begin{array}{cl}
0  \;\;\;\;\;\;\;\;\;\; &,n=0, \\
\nonumber \\
\sqrt{n} \phi_{n-1} \;\;\; &,n=1,2, \cdots,
\end{array}
\right. \nonumber \\
B_{\bm{f}} \phi_{n}
&=& \sqrt{n+1} \phi_{n+1} \;\;\;\;\;\;\;\;\; ,n=0,1, \cdots , \nonumber \\
\nonumber \\
A_{\bm{f}}^{\dagger} \psi_{n}
&=& \sqrt{n+1} \psi_{n+1} \;\;\;\;\;\;\;\;\; ,n=0,1, \cdots , \nonumber \\
B_{\bm{f}}^{\dagger} \psi_{n}
&=& \left\{
\begin{array}{cl}
0  \;\;\;\;\;\;\;\;\;\; &,n=0, \\
\nonumber \\
\sqrt{n} \psi_{n-1} \;\;\; &,n=1,2, \cdots,
\end{array}
\right. \nonumber \\
\nonumber \\
N_{\bm{f}}\phi_{n} 
&=& n \phi_{n} \;\;\;\;\;\;\;\;\;\;\;\;\;\;\;\;\;\;\;\;\;\; \; ,n=0,1, \cdots , \nonumber \\
N_{\bm{f}}^{\dagger} \psi_{n}
&=& n \psi_{n} \;\;\;\;\;\;\;\;\;\;\;\;\;\;\;\;\;\;\;\;\;\; \; ,n=0,1, \cdots , \nonumber 
\end{eqnarray}
and furthermore
\begin{eqnarray}
A_{\bm{f}}B_{\bm{f}} - B_{\bm{f}}A_{\bm{f}} \subset I \;\;\;{\rm  and} \;\;\;
B_{\bm{f}}^{\dagger} A_{\bm{f}}^{\dagger} -A_{\bm{f}}^{\dagger}B_{\bm{f}}^{\dagger} \subset I . \nonumber
\end{eqnarray}
Thus, $A_{\bm{f}}$ and $B_{\bm{f}}$ are lowering and raising operators for$\{ \phi_{n} \}$, respectively, and $A_{\bm{f}}^{\dagger}$ and $B_{\bm{f}}^{\dagger}$ are raising and lowering operators for $\{ \psi_{n} \}$, respectively, and $N_{\bm{f}}$ and $N_{\bm{f}}^{\dagger}$ are number operators for $\{ \phi_{n} \}$ and $\{ \psi_{n} \}$, respectively. These operators connect with ${\it quasi}$-${\it hermitian \; quantum \; mechanics}$ and its relatives. Many researchers have investigated such operators mathematically. \cite{h-t, b-i-t}. Hereafter, we shall consider the connection with ${\it pseudo}$-${\it bosons}$, where in the recent literature many researchers have investigated. \cite{bagarello13, bagarello10, bagarello11, mostafazadeh, d-t} This paper clarifies and simplifies the mathematical structure of this framework minimized the required assumptions.

This article is organized as follows. In Section 2, we define and study the notions of biorthogonal and regular pairs. By using the notions, we introduce general theories of a regular biorthogonal pair $( \{ \phi_{n} \}, \; \{ \psi_{n} \} )$. In Section 3, we define and study the lowering operators $A_{\bm{f}}$ and $B_{\bm{f}}^{\dagger}$, the raising operators $B_{\bm{f}}$ and $A_{\bm{f}}^{\dagger}$, the number operators $N_{\bm{f}}$ and $N_{\bm{f}}^{\dagger}$ determined by the non-singular positive self-adjoint operator $T_{\bm{f}}$. By using the notions, we introduce general theories of these operators.
\section{General theory of generalized Riesz bases}
In this section, we define the notion of biorthogonal and regular pairs and introduce its general theory. In particular, we investigate the relationship between a regular biorthogonal pair $( \{ \phi_{n} \} , \{ \psi_{n} \} )$ and the notions of Riesz bases and semi-Riesz bases.\\
\par
{\it Definition 2.1.} {\it Sequences $\{ \phi_{n} \}$ and $\{ \psi_{n} \}$ in a Hilbert space ${\cal H}$ are said to be biorthogonal if $( \phi_{n} | \psi_{m} ) = \delta_{nm}$, $n,\; m=0,1, \cdots .$}\\\\
It is easily shown that both $\{ \phi_{n} \}$ and $\{ \psi_{n} \}$ are linearly independent. Furthermore, $D_{\phi} \equiv Span \; \{ \phi_{n} \}$ (or $D_{\psi} \equiv Span \; \{ \psi_{n} \}$) is not necessarily dense in ${\cal H}$. Indeed, we put $\{ \phi_{n} \equiv e_{n+1}+e_{1}, \; n=1,2, \cdots \}$ and $\{ \psi_{n} \equiv e_{n+1}, \; n=1,2, \cdots \}$, where $\{ e_{n} \}$ is an ONB in ${\cal H}$. Then $\{ \phi_{n} \}$ and $\{ \psi_{n} \}$ are biorthogonal and $D_{\phi}$ is dense in ${\cal H}$, but $D_{\psi}$ is not dense in ${\cal H}$. This example is given in Ref. \cite{bagarello13}. In more general, let $\{ e_{n} \}$ be an orthonormal system such that $Span \; \{ e_{n} \}$ is not dense in ${\cal H}$ and $T$ be a bounded operator on ${\cal H}$ with bounded inverse. We put 
\begin{eqnarray}
\phi_{n}=Te_{n} \;\;\; {\rm and} \;\;\; \psi_{n}=(T^{-1})^{\ast} e_{n}, \;\;\; n=0,1, \cdots. \nonumber
\end{eqnarray}
Then it is easily shown that $\{ \phi_{n} \}$ and $\{ \psi_{n} \}$ are biorthogonal, but both $D_{\phi}$ and $D_{\psi}$ are not dense in ${\cal H}$. Indeed, take an arbitrary $y \neq 0 \in \{ e_{n} \}^{\perp}$. Then since $T^{-1}y \neq 0$ and $( \phi_{n} | T^{-1}y)=(e_{n} |y)=0$ for $n=0,1, \cdots$, we have $T^{-1}y \in \{ \phi_{n} \}^{\perp}$. Hence $D_{\phi}$ is not dense in ${\cal H}$. Similarly, we have $Ty \neq 0 \in \{ \psi_{n} \}^{\perp}$. Hence, $D_{\psi}$ is not dense in ${\cal H}$. In Theorem 2.3, we shall give necessary and sufficient conditions for biorthogonal sequences $\{ \phi_{n} \}$ and $\{ \psi_{n} \}$ under which $D_{\phi}$ and $D_{\psi}$ are dense in ${\cal H}$. Thus, various cases for biorthogonal sequences arise.\\
\par
{\it Definition 2.2.} {\it A pair $( \{ \phi_{n} \}, \; \{ \psi_{n} \} )$ of biorthogonal sequences  is said to be regular if both $D_{\phi}$ and $D_{\psi}$ are dense in ${\cal H}$.}\\\\
Let $\{ \phi_{n} \}$ and $\{ \psi_{n} \}$ be biorthogonal. For any ONB $\bm{e} \equiv \{ e_{n} \}$ in ${\cal H}$, we put
\begin{equation}
T_{\bm{e}} \left( \sum_{k=0}^{n} \alpha_{k} e_{k} \right) = \sum_{k=0}^{n} \alpha_{k} \phi_{k}, \tag{2.1} 
\end{equation}
\begin{equation}
K_{\bm{e}} \left( \sum_{k=0}^{n} \alpha_{k} e_{k} \right) = \sum_{k=0}^{n} \alpha_{k} \psi_{k}. \tag{2.2}
\end{equation}
for $\sum_{k=0}^{n} \alpha_{k}e_{k} \in D_{\bm{e}} \equiv Span \; \{ e_{n} \}$. Then $T_{\bm{e}}$ and $K_{\bm{e}}$ are densely defined linear operators in ${\cal H}$ and we have
\begin{align}
\left( T_{\bm{e}} \left( \sum_{k=0}^{n} \alpha_{k} e_{k} \right) \left| \; K_{\bm{e}} \left( \sum_{j=0}^{m} \beta_{j} e_{j} \right) \right. \right) 
&= \sum_{k=0}^{n} \sum_{j=0}^{m} \alpha_{k} \bar{\beta}_{j} ( \phi_{k} | \psi_{j}) \nonumber \\
&= \sum_{k=0}^{n} \sum_{j=0}^{m} \alpha_{k} \bar{\beta}_{j} \delta_{kj} \nonumber \\
&= \left\{
\begin{array}{cl}
& \sum_{k=0}^{n}  \alpha_{k} \bar{\beta}_{k}, \;\;\; n \leq m  \\\\
& \sum_{k=0}^{m} \alpha_{k} \bar{\beta}_{k}, \;\;\; n>m \\
\end{array}
\right. \nonumber \\
&= \left( \sum_{k=0}^{n} \alpha_{k}e_{k} \left| \; \sum_{j=0}^{m} \beta_{j}e_{j} \right) \right. \tag{2.3}
\end{align}
for any $\sum_{k=0}^{n} \alpha_{k}e_{k}$, $\sum_{j=0}^{m} \beta_{j}e_{j} \; \in D_{\bm{e}}$, which implies that
\begin{equation}
K_{\bm{e}}^{\ast}T_{\bm{e}} =I \;\;\; {\rm on} \;\;\; D_{\bm{e}}\; . \tag{2.4}
\end{equation}
But, $T_{\bm{e}}$ and $K_{\bm{e}}$ are not necessarily closable. In next Theorem 2.3, we shall show that $T_{\bm{e}}$ and $K_{\bm{e}}$ are closable if and only if $( \{ \phi_{n} \}, \; \{ \psi_{n} \} )$ is regular. \\
\par
{\it Theorem 2.3.} {\it Let $\{ \phi_{n} \}$ and $\{ \psi_{n} \}$ be sequences in ${\cal H}$. Then the following statements are equivalent.\\
\hspace{3mm} (i) $( \{ \phi_{n} \}, \; \{ \psi_{n} \} )$ is a regular biorthogonal pair.\\
\hspace{3mm} (ii) $\{ \phi_{n} \}$ and $\{ \psi_{n} \}$ are biorthogonal, and $T_{\bm{e}}$ and $K_{\bm{e}}$ are closable for any ONB $\bm{e} \equiv \{ e_{n} \}$ in ${\cal H}$.\\
\hspace{3mm} (iii) For any ONB $\bm{e}= \{e_{n} \}$ in ${\cal H}$ there exists a densely defined closed linear operator $T$ in ${\cal H}$ with densely defined inverse such that $ \bm{e} \subset D(T) \cap D( (T^{-1} )^{\ast}), \; Te_{n}= \phi_{n}$ and $(T^{-1})^{\ast} e_{n}=\psi_{n}, \; n=0,1, \cdots .$\\
\hspace{3mm} (iv) There exists an ONB $\bm{f} = \{ f_{n} \}$ in ${\cal H}$ such that $T_{\bm{f}}$ is a non-singular positive essentially self-adjoint operator in ${\cal H}$, and it is unique in the following sense: if $\bm{g}$ is an ONB in ${\cal H}$ such that $T_{\bm{g}}$ is a non-singular positive essentially self-adjoint operator in ${\cal H}$, then $\bm{g}=\bm{f}$. }\\\\
If this holds, then for any ONB $\bm{e}$, the closure $\bar{T}_{\bm{e}}$ of $T_{\bm{e}}$ (for simplicity denoted by the same $T_{\bm{e}}$ after the proof of Lemma 2.4) is the minimum in closed operators $T$ satisfying condition in (iii). \\
\par
{\it Proof.} (i)$\Rightarrow$(ii) Since $D(T_{\bm{e}}^{\ast}) \supset D_{\psi}$ and $D(K_{\bm{e}}^{\ast}) \supset D_{\phi}$ by (2.3), it follows that $D(T_{\bm{e}}^{\ast})$ and $D(K_{\bm{e}}^{\ast})$ are dense in ${\cal H}$, equivalently, $T_{\bm{e}}$ and $K_{\bm{e}}$ are closable.\\
(ii)$\Rightarrow$(iii) We show that $\bar{T}_{\bm{e}}$ satisfies conditions in (iii).\\ Indeed, for any $x \in D(\bar{T}_{\bm{e}})$ there exists a sequence $ \{ x_{n} \} $ in $D(T_{\bm{e}}) = Span \; \bm{e}$ such that $\lim_{n \rightarrow \infty} x_{n}=x$ and $\lim_{n \rightarrow \infty} T_{\bm{e}}x_{n}=\bar{T}_{\bm{e}} x$. By (2.4), $\lim_{n \rightarrow \infty} K_{\bm{e}}^{\ast}T_{\bm{e}}x_{n}=\lim_{n \rightarrow \infty} x_{n}=x$. Hence, we have
\begin{eqnarray}
\bar{T}_{\bm{e}}x \in D(K_{\bm{e}}^{\ast}) \;\; {\rm and} \;\; K_{\bm{e}}^{\ast} \bar{T}_{\bm{e}}x=x, \nonumber
\end{eqnarray}
which means that $\bar{T}_{\bm{e}}^{-1}=K_{\bm{e}}^{\ast}$. Since $K_{\bm{e}}$ is closable, we have
\begin{equation}
\bar{K}_{\bm{e}}=K_{\bm{e}}^{\ast \ast} = ( \bar{T}_{\bm{e}}^{-1})^{\ast}. \tag{2.5}
\end{equation}
Thus $\bar{T}_{\bm{e}}$ is a densely defined closed operator with densely defined inverse such that $\bm{e} \subset D(\bar{T}_{\bm{e}}) \cap D((\bar{T}_{\bm{e}}^{-1})^{\ast})$, $\bar{T}_{\bm{e}}e_{n}=\phi_{n}$ and $(\bar{T}_{\bm{e}}^{-1})^{\ast}e_{n}=\psi_{n}, \; n=0,1, \cdots .$\\
(iii)$\Rightarrow$(iv) For any ONB $\bm{e}= \{ e_{n} \}$ in ${\cal H}$, $T_{\bm{e}} \subset T$ and $K_{\bm{e}} \subset (T^{-1})^{\ast}$. Hence $T_{\bm{e}}$ and $K_{\bm{e}}$ are closable. Let $T_{\bm{e}}^{\ast}=U|T_{\bm{e}}^{\ast}|$ be the polar decomposition of $T_{\bm{e}}^{\ast}$. Since $(T_{\bm{e}}^{\ast})^{-1}=\bar{K}_{\bm{e}}$ by (2.5), it follows that $T_{\bm{e}}^{\ast}$ has a densely defined inverse, which implies that $U$ is a unitary operator. We here put $\bm{f}= \{ U^{\ast} e_{n} \}$. Then it is shown that $\bm{f}$ is an ONB in ${\cal H}$ and $\bar{T}_{\bm{f}}=|T_{\bm{e}}^{\ast}|$. In detail, this statement is proved in Ref. \cite{h-t}. Furthermore, the uniqueness is shown in Lemma 2.4.\\
(iv)$\Rightarrow$(i) Since
\begin{eqnarray}
(\phi_{n} | \psi_{m})
&=& (T_{\bm{f}} f_{n}|T_{\bm{f}}^{-1}f_{m}) \nonumber \\
&=& (f_{n}|f_{m}) \nonumber \\
&=& \delta_{nm}, \;\;\;\; n,m=0,1, \cdots , \nonumber
\end{eqnarray}
we have that $\{ \phi_{n} \}$ and $\{ \psi_{n} \}$ are biorthogonal. Next, we show that $D_{\phi}$ and $D_{\psi}$ are dense in ${\cal H}$. Take an arbitrary $y \in \{ \phi_{n} \}^{\perp}$. Then,
\begin{eqnarray}
0
&=& ( y| \; \sum_{k=0}^{n} \alpha_{k} \phi_{k} ) \nonumber \\
&=& \left(y\; | \; T_{\bm{f}} \left( \sum_{k=0}^{n} \alpha_{k} f_{k} \right) \right) \nonumber 
\end{eqnarray}
for any $x= \sum_{k=0}^{n} \alpha_{k} \phi_{k} \in D_{\phi}$. Hence we have
\begin{eqnarray}
y \in D(T_{\bm{f}}^{\ast})=D(\bar{T}_{\bm{f}}) \;\; {\rm and} \;\; \bar{T}_{\bm{f}}y=0. \nonumber
\end{eqnarray}
Since $T_{\bm{f}}$ has inverse, we have $y=0$. Therefore $D_{\phi}$ is dense in ${\cal H}$. It is similarly shown that $D_{\psi}$ is dense in ${\cal H}$. This completes the proof. \\\\
Let $(\{ \phi_{n} \}, \{ \psi_{n} \})$ be a regular biorthogonal pair. We investigate the relevance of these operators $T_{\bm{e}}$ and $T_{\bm{g}}$, defined by ONB  $\bm{e}$ and  ONB $\bm{g}$ in ${\cal H}$. We define a unitary operator $U_{\bm{e}, \bm{g}}$ by
\begin{eqnarray}
U_{\bm{e}, \bm{g}} \left( \sum_{k=0}^{n} \alpha_{k}e_{k} \right) = \sum_{k=0}^{n} \alpha_{k}g_{k}. \nonumber
\end{eqnarray}
Then we have the following\\
\par
{\it Lemma 2.4.} {\it The following statements hold.\\
\hspace{3mm} (1) $\bar{T}_{\bm{e}}=\bar{T}_{\bm{g}}U_{\bm{e}, \bm{g}}$.\\
\hspace{3mm} (2) $\bar{T}_{\bm{e}}T_{\bm{e}}^{\ast}=\bar{T}_{\bm{g}}T_{\bm{g}}^{\ast}$. Hence, $|T_{\bm{e}}^{\ast}|=|T_{\bm{g}}^{\ast}|. $}\\
\par
{\it Proof.} (1) Take an arbitrary $x \in D(\bar{T}_{\bm{e}})$. Then there exists a sequence $\{ x_{n} \}$ in $Span \; \bm{e}$ such that $\lim_{n \rightarrow \infty} x_{n} = x$ and $\lim_{n \rightarrow \infty} T_{\bm{e}} x_{n} =\bar{T}_{\bm{e}} x$. Since,
\begin{eqnarray}
T_{\bm{g}} U_{\bm{e}, \bm{f}} \left( \sum_{k=0}^{n} \alpha_{k}e_{k} \right) 
&=&  T_{\bm{g}} \left( \sum_{k=0}^{n} \alpha_{k}g_{k} \right) \nonumber \\
&=&   \sum_{k=0}^{n} \alpha_{k}\phi_{k}  \nonumber \\
&=&  T_{\bm{e}} \left( \sum_{k=0}^{n} \alpha_{k}e_{k} \right) \nonumber 
\end{eqnarray}
for any $\sum_{k=0}^{n} \alpha_{k} e_{k} \in D_{\bm{e}}$, it follows that
\begin{eqnarray}
\lim_{n \rightarrow \infty} U_{\bm{e}, \bm{g}}x_{n}
&=& U_{\bm{e}, \bm{g}}x \;\;\; {\rm and} \nonumber \\
\lim_{n \rightarrow \infty} T_{\bm{g}} U_{\bm{e}, \bm{g}}x_{n}
&=& \lim_{n \rightarrow \infty} T_{\bm{e}} x_{n} \nonumber \\
&=& \bar{T}_{\bm{e}} x . \nonumber
\end{eqnarray}
Hence, we have
\begin{eqnarray}
U_{\bm{e}, \bm{g}} x \in D(\bar{T}_{\bm{g}}) \;\;\;{\rm and} \;\;\; \bar{T}_{\bm{g}}U_{\bm{e}, \bm{g}}x=\bar{T}_{\bm{e}}x . \nonumber
\end{eqnarray}
Thus, we have
\begin{eqnarray}
\bar{T}_{\bm{e}} \subset \bar{T}_{\bm{g}} U_{\bm{e}, \bm{g}}. \nonumber
\end{eqnarray}
Similarly, we have
\begin{eqnarray}
\bar{T}_{\bm{g}} \subset \bar{T}_{\bm{e}} U_{\bm{g},\bm{e}} = \bar{T}_{\bm{e}} (U_{\bm{e}, \bm{g}})^{\ast}. \nonumber
\end{eqnarray}
Hence we have
\begin{eqnarray}
\bar{T}_{\bm{g}} U_{\bm{e},\bm{g}} \subset \bar{T}_{\bm{e}}. \nonumber
\end{eqnarray}
Thus we have
\begin{eqnarray}
\bar{T}_{\bm{e}}=\bar{T}_{\bm{g}} U_{\bm{e},\bm{g}}. \nonumber 
\end{eqnarray}
\\
(2) By (1), we have $T_{\bm{e}}^{\ast} \supset U_{\bm{e},\bm{g}}^{\ast} T_{\bm{g}}^{\ast}$. Hence it follow that
\begin{eqnarray}
\bar{T}_{\bm{e}} T_{\bm{e}}^{\ast} \supset \bar{T}_{\bm{g}} U_{\bm{e},\bm{g}}U_{\bm{e},\bm{g}}^{\ast} T_{\bm{g}}^{\ast} = \bar{T}_{\bm{g}}T_{\bm{g}}^{\ast} . \nonumber
\end{eqnarray}
Furthermore, by replacing the $\bm{e}$ and $\bm{g}$ we have
\begin{eqnarray}
\bar{T}_{\bm{g}}T_{\bm{g}}^{\ast} \supset \bar{T}_{\bm{e}}T_{\bm{e}}^{\ast}. \nonumber
\end{eqnarray}
From the above, 
\begin{eqnarray}
\bar{T}_{\bm{g}}T_{\bm{g}}^{\ast} = \bar{T}_{\bm{e}}T_{\bm{e}}^{\ast}. \nonumber
\end{eqnarray}
This completes the proof.
\\\\
By Lemma 2.4, we can show the uniqueness of $\{ f_{n} \}$ in Theorem 2.3. (iv), and so the proof of Theorem 2.3 completes. Hereafter we denote the closure of $T_{\bm{e}}$ by the same $T_{\bm{e}}$ excluding the case when lead to confusion.\\ 
Next we define and study the notions of Riesz bases and semi-Riesz bases. And we investigate the relationship these notions and regular biorthogonal pairs.\\
\par
{\it Definition 2.5.} {\it Let $\{ \phi_{n} \}$ and $\{ \psi_{n} \}$ be sequences in ${\cal H}$. $( \{ \phi_{n} \} , \{ \psi_{n} \} )$ is said to be a pair of Riesz bases if\\
\hspace{3mm} (i) $( \{ \phi_{n} \} , \{ \psi_{n} \} )$ is a regular biorthogonal pair.\\
\hspace{3mm} (ii) There exists an ONB $ \bm{e} = \{ e_{n} \}$ in ${\cal H}$ such that both $T_{\bm{e}}$ and $(T_{\bm{e}}^{-1})^{\ast}$ are bounded.}\\
\par 
{\it Definition 2.6.}  {\it Let $\{ \phi_{n} \}$ and $\{ \psi_{n} \}$ be sequences in ${\cal H}$. $( \{ \phi_{n} \} , \{ \psi_{n} \} )$ is said to be a pair of semi-Riesz bases if\\
\hspace{3mm} (i) $( \{ \phi_{n} \} , \{ \psi_{n} \} )$ is a regular biorthogonal pair.\\
\hspace{3mm} (ii) There exists an ONB $\bm{e} = \{ e_{n} \}$ in ${\cal H}$ such that either $T_{\bm{e}}$ or $(T_{\bm{e}}^{-1})^{\ast}$ are bounded.}\\\\
By Lemma 2.4, we have the following.\\
\par
{\it Lemma 2.7.} {\it If $T_{\bm{e}}$ (resp. $T_{\bm{e}}^{-1}$) is bounded for some ONB $\bm{e}$, then $T_{\bm{g}}$ (resp. $T_{\bm{g}}^{-1}$) is bounded for any ONB $\bm{g}$ in ${\cal H}$.}\\\\
This means that the notions of Riesz bases and semi-Riesz bases do not depend on methods of taking ONB. In other word, the following holds.\\
\par
{\it Lemma 2.8} {\it $( \{ \phi_{n} \}, \{ \psi_{n} \})$ is a pair of regular biorthogonal sequences. Then the folloing statements are equivalent:\\
\hspace{3mm} (i) $( \{ \phi_{n} \}, \{ \psi_{n} \})$ is a pair of Riesz bases (resp. semi-Riesz bases) for some ONB $\bm{e}= \{ e_{n} \}$ in ${\cal H}$.\\
\hspace{3mm} (ii) $( \{ \phi_{n} \}, \{ \psi_{n} \})$ is a pair of Riesz bases (resp. semi-Riesz bases) for any ONB $\bm{g}= \{ g_{n} \}$ in ${\cal H}$.}\\\\
We proceed studies of regular biorthogonal pairs. Let $( \{ \phi_{n} \}, \{ \psi_{n} \})$ be a regular biorthogonal pair. Then we define two operators $S_{\phi}$ and $S_{\psi}$ by
\begin{eqnarray}
S_{\phi} &=& \sum_{n=0}^{\infty} \phi_{n} \otimes \bar{\phi}_{n} \nonumber 
\end{eqnarray}
and
\begin{eqnarray}
S_{\psi} &=& \sum_{n=0}^{\infty} \psi_{n} \otimes \bar{\psi}_{n}, \nonumber
\end{eqnarray}
where the tensor $x \otimes \bar{y}$ of elements $x, \; y$ of ${\cal H}$ is defined by
\begin{eqnarray}
(x \otimes \bar{y})\xi =( \xi | y)x, \;\;\; \xi \in {\cal H}. \nonumber
\end{eqnarray}
In detail, the operators $S_{\phi}$ and $S_{\psi}$ are defined as follows:
\begin{equation}
\left\{
\begin{array}{cl}
 D(S_{\phi})
&= \ \left\{ x \in {\cal H} \ ; \  \lim_{n \rightarrow \infty} \sum_{k=0}^{n} (x | \phi_{k}) \phi_{k} \ {\rm exists} \ {\rm in} \ {\cal H} \right\}  \\\\
S_{\phi}x
&= \  \sum_{n=0}^{\infty} (x| \phi_{n}) \phi_{n}, \;\;\; x \in D(S_{\phi}) \\
\end{array}
\right. \nonumber \\ 
\end{equation}
\begin{equation}
\left\{
\begin{array}{cl}
 D(S_{\psi})
&= \ \left\{ x \in {\cal H} \ ; \  \lim_{n \rightarrow \infty} \sum_{k=0}^{n} (x | \psi_{k}) \psi_{k} \ {\rm exists} \ {\rm in} \ {\cal H} \right\}  \\\\
S_{\psi}x
&= \  \sum_{n=0}^{\infty} (x| \psi_{n}) \psi_{n}, \;\;\; x \in D(S_{\psi}) .\\
\end{array}
\right. \nonumber \\ 
\end{equation}
We investigate the properties of $S_{\phi}$ and $S_{\psi}$ and the relationships between $S_{\phi}$, $S_{\psi}$ and $T_{\bm{e}}T_{\bm{e}}^{\ast}$, $(T_{\bm{e}}^{-1})^{\ast}T_{\bm{e}}^{-1}$, respectively. It is easily shown that
\begin{eqnarray}
D(S_{\phi}) &\supset& D_{\psi} \;\;\; {\rm and} \;\;\; S_{\phi}\psi_{n}=\phi_{n}, \;\;\; n=0,1, \cdots , \nonumber \\
D(S_{\psi}) &\supset& D_{\phi} \;\;\; {\rm and} \;\;\; S_{\psi}\phi_{n}=\psi_{n}, \;\;\; n=0,1, \cdots . \nonumber
\end{eqnarray}
Hence, $S_{\phi}$ and $S_{\psi}$ are densely defined linear operators in ${\cal H}$ satisfying
\begin{eqnarray}
S_{\phi}S_{\psi} &=& 1 \;\;\; {\rm on} \;\;\; D_{\phi}, \nonumber \\
S_{\psi}S_{\phi} &=& 1 \;\;\; {\rm on} \;\;\; D_{\psi}. \nonumber
\end{eqnarray}
Since $D_{\phi}$ and $D_{\psi}$ are dense in ${\cal H}$, we have
\begin{eqnarray}
\overline{S_{\phi}S_{\psi}} = \overline{S_{\psi}S_{\phi}} =1. \nonumber
\end{eqnarray}
Furthermore, it is easily shown that $S_{\phi}$ and $S_{\psi}$ are positive and symmetric operators in ${\cal H}$.\\
\par
{\it Lemma 2.9.} {\it The following statements holds.\\
\hspace{3mm} (1) If $( \{ \phi_{n} \} , \{ \psi_{n} \})$ is a regular biorthogonal pair, then for any ONB $\bm{e}= \{ e_{n} \}$ in ${\cal H}$
\begin{eqnarray}
S_{\phi} &=&T_{\bm{e}}T_{\bm{e}}^{\ast} \;\;\;\;\;\;\;\;\;\;\;\; {\rm on} \;\;\; D(S_{\phi}) \cap D(T_{\bm{e}}^{\ast}), \nonumber \\
S_{\psi} &=&(T_{\bm{e}}^{-1})^{\ast} T_{\bm{e}}^{-1} \;\;\; {\rm on} \;\;\; D(S_{\psi}) \cap D(T_{\bm{e}}^{-1}). \nonumber
\end{eqnarray}
\hspace{3mm} (2) Let $( \{ \phi_{n} \} , \{ \psi_{n} \})$ be a pair of semi-Riesz bases and $\bm{e}$ be an ONB in ${\cal H}$. In case that $T_{\bm{e}}$ is unbounded and $T_{\bm{e}}^{-1}$ is bounded,
\begin{eqnarray}
S_{\phi} \subset T_{\bm{e}}T_{\bm{e}}^{\ast} \;\;\; {\rm and} \;\;\; S_{\psi} =(T_{\bm{e}}^{-1})^{\ast} T_{\bm{e}}^{-1}, \nonumber
\end{eqnarray}
and in case that $T_{\bm{e}}$ is bounded and $T_{\bm{e}}^{-1}$ is unbounded,
\begin{eqnarray}
S_{\phi} = T_{\bm{e}}T_{\bm{e}}^{\ast} \;\;\; {\rm and} \;\;\; S_{\psi}  \subset (T_{\bm{e}}^{-1})^{\ast} T_{\bm{e}}^{-1}. \nonumber
\end{eqnarray}
\hspace{3mm} (3) If $( \{ \phi_{n} \} , \{ \psi_{n} \})$ is a pair of Riesz bases, then
\begin{eqnarray}
S_{\phi} = T_{\bm{e}}T_{\bm{e}}^{\ast} \;\;\; {\rm and} \;\;\; S_{\psi}  = (T_{\bm{e}}^{-1})^{\ast} T_{\bm{e}}^{-1}. \nonumber
\end{eqnarray}}\\
\par
{\it Proof.} (1) Take an arbitrary $x \in D(S_{\phi}) \cap D(T_{\bm{e}}^{\ast})$. Since $x \in D(T_{\bm{e}}^{\ast})$, we have
\begin{eqnarray}
 \left( \sum_{k=0}^{n} \phi_{k} \otimes \bar{\phi}_{k} \right) x
&=& \sum_{k=0}^{n} ( x | \phi_{k}) \phi_{k} \nonumber \\
&=& \sum_{k=0}^{n} ( x| T_{\bm{e}}e_{k} ) T_{\bm{e}}e_{k} \nonumber \\
&=& T_{\bm{e}} \left( \sum_{k=0}^{n} (T_{\bm{e}}^{\ast} x | e_{k}) e_{k} \right) , \nonumber
\end{eqnarray}
which implies that
\begin{eqnarray}
\lim_{n \rightarrow \infty} \sum_{k=0}^{n} (T_{\bm{e}}^{\ast} x |e_{k})e_{k}
= T_{\bm{e}}^{\ast}x, \nonumber
\end{eqnarray}
and 
\begin{eqnarray}
\lim_{n \rightarrow \infty} T_{\bm{e}} \left( \sum_{k=0}^{n} (T_{\bm{e}}x|e_{k}) e_{k} \right)
&=& \lim_{n \rightarrow \infty} \left( \sum_{k=0}^{n} \phi_{k} \otimes \bar{\phi}_{k} \right) x \nonumber \\
&=& S_{\phi} x. \nonumber
\end{eqnarray}
Thus we have
\begin{eqnarray}
T_{\bm{e}}^{\ast}x \in D(T_{\bm{e}}) \;\;\; {\rm and} \;\;\; T_{\bm{e}}T_{\bm{e}}^{\ast}x= S_{\phi}x. \nonumber
\end{eqnarray}
From the above,
\begin{eqnarray}
S_{\phi}=T_{\bm{e}}T_{\bm{e}}^{\ast} \;\;\; {\rm on} \;\;\; D(S_{\phi}) \cap D(T_{\bm{e}}^{\ast}). \nonumber
\end{eqnarray}
The statement for $S_{\psi}$ is proved in similar way.\\
(2) We assume that $T_{\bm{e}}$ is unbounded and $T_{\bm{e}}^{-1}$ is bounded. Take an arbitrary $x \in D(S_{\phi})$. Then we have
\begin{eqnarray}
\lim_{n \rightarrow \infty}T_{\bm{e}} \left( \sum_{k=0}^{n} ( x| T_{\bm{e}} e_{k})e_{k} \right) 
&=& \lim_{n \rightarrow \infty} \sum_{k=0}^{n} ( x| \phi_{k}) \phi_{k} \nonumber \\
&=& S_{\phi} x. \nonumber 
\end{eqnarray}
Since $T_{\bm{e}}^{-1}$ is bounded, it follows that
\begin{eqnarray}
\lim_{n \rightarrow \infty} \sum_{k=0}^{n} ( x| T_{\bm{e}} e_{k})e_{k} 
= T_{\bm{e}}^{-1} S_{\phi} x . \nonumber
\end{eqnarray}
Thus, we have
\begin{equation}
\sum_{k=0}^{n} |(x| T_{\bm{e}} e_{k} )|^{2} < \infty . \tag{2.6}
\end{equation}
Furthermore, take an arbitrary $y= \sum_{k=0}^{\infty} (y|e_{k})e_{k} \in D(T_{\bm{e}})  $. Then there exists a sequence 
\begin{eqnarray}
\left\{ y_{n} \equiv \sum_{k=0}^{l_{n}} (y_{n} |e_{k})e_{k} \right\} \;\;\; {\rm in} \;\;\; D_{\bm{e}} \nonumber
\end{eqnarray}
such that 
\begin{eqnarray}
\lim_{n \rightarrow \infty} y_{n} &=& y, \nonumber \\
\lim_{n \rightarrow \infty} T_{\bm{e}}y_{n}
&=& \lim_{n \rightarrow \infty} \sum_{k=0}^{l_{n}} (y_{n} |e_{k}) T_{\bm{e}} e_{k} \nonumber \\
&=& T_{\bm{e}} y. \nonumber 
\end{eqnarray}
Then it follows from (2.6) and Schwartz's inequality that
\begin{eqnarray}
|(T_{\bm{e}} y|x)|
&=& \lim_{n \rightarrow \infty} |(T_{\bm{e}} y_{n} |x)| \nonumber \\
&=& \lim_{n \rightarrow \infty}\left| \sum_{k=0}^{l_{n}} (y_{n}|e_{k})(T_{\bm{e}}e_{k}| x) \right| \nonumber \\
& \leq& \lim_{n \rightarrow \infty} \left( \sum_{k=0}^{l_{n}} |( y_{n}|e_{k})|^{2} \right)^{\frac{1}{2}} \left( \sum_{k=0}^{l_{n}} |( T_{\bm{e}}e_{k}| x)|^{2} \right)^{\frac{1}{2}} \nonumber \\
&=& \| y \| \left( \sum_{k=0}^{\infty} |( T_{\bm{e}}e_{k}| x)|^{2} \right)^{\frac{1}{2}}, \nonumber 
\end{eqnarray}
which implies $x \in D(T_{\bm{e}}^{\ast})$ and $D(S_{\phi}) \subset D(T_{\bm{e}}^{\ast})$. By (1), we have
\begin{eqnarray}
S_{\phi}=T_{\bm{e}}T_{\bm{e}}^{\ast} \;\;\; {\rm on} \;\;\; D(S_{\phi}), \nonumber
\end{eqnarray}
which means that
\begin{eqnarray}
S_{\phi} \subset T_{\bm{e}}T_{\bm{e}}^{\ast}. \nonumber
\end{eqnarray}
Furthermore, take an arbitrary $x \in {\cal H}$. Since $T_{\bm{e}}^{-1}$ is bounded, it follows that
\begin{eqnarray}
\lim_{n \rightarrow \infty} \sum_{k=0}^{n} ( x| \psi_{k})\psi_{k}
&=& \lim_{n \rightarrow \infty} (T_{\bm{e}}^{-1})^{\ast} \left( \sum_{k=0}^{n} (T_{\bm{e}}^{-1} x|e_{k})e_{k} \right) \nonumber \\
&=& (T_{\bm{e}}^{-1})^{\ast} \left( \sum_{k=0}^{\infty} (T_{\bm{e}}^{-1} x|e_{k})e_{k} \right) \nonumber \\
&=&  (T_{\bm{e}}^{-1})^{\ast}T_{\bm{e}}^{-1} x,  \nonumber
\end{eqnarray}
which implies that $x \in D(S_{\psi})$ and $S_{\psi}=( T_{\bm{e}}^{-1})^{\ast} T_{\bm{e}}^{-1}.$ Thus we have
\begin{eqnarray}
S_{\psi}=( T_{\bm{e}}^{-1})^{\ast} T_{\bm{e}}^{-1}. \nonumber
\end{eqnarray}
(3) This follows from (1) and (2). This completes the proof.\\\\
By the proofs of Lemma 2.9, we have the following\\
\par
{\it Remark.} {\it The following statements are equivalent:\\
\hspace{3mm} (i) $S_{\phi} \subset T_{\bm{e}}T_{\bm{e}}^{\ast}.$ \\
\hspace{3mm} (ii) $D(S_{\phi}) \subset D(T_{\bm{e}}^{\ast}). $\\
\hspace{3mm} (iii) $\sum_{k=0}^{\infty} |( x | \phi_{k})|^{2} < \infty, \;\;\; {\rm for \; any} \; x \in D(S_{\phi}).$\\
\hspace{3mm} Similarly, the following statements are equivalent,\\
\hspace{3mm} (i)' $S_{\psi} \subset (T_{\bm{e}}^{-1})^{\ast} T_{\bm{e}}^{-1}$.\\
\hspace{3mm} (ii)' $D(S_{\psi}) \subset D(T_{\bm{e}}^{-1}).$ \\
\hspace{3mm} (iii)' $\sum_{k=0}^{\infty} |( x| \psi_{k})|^{2} < \infty, \;\;\; {\rm for \; any} \; x \in D(S_{\psi}).$}\\\\
We characterize the notions of Riesz bases and semi-Riesz bases.\\
\par
{\it Proposition 2.10.} {\it Let $( \{ \phi_{n} \}, \{ \psi_{n} \})$ be a regular biorthogonal pair. Then the following statements are equivalent:\\
\hspace{3mm} (i) $( \{ \phi_{n} \}, \{ \psi_{n} \})$ is a pair of Riesz bases.\\
\hspace{3mm} (ii) $\{ \phi_{n} \}$ and $\{ \psi_{n} \}$ are Bessel sequences, that is, there exists positive constants $r_{\phi}$ and \\
\hspace{9mm} $r_{\psi}$ such that
\begin{eqnarray}
\sum_{k=0}^{\infty} |( x|\phi_{n})|^{2} &\leq& r_{\phi} \| x \|^{2} , \nonumber 
\end{eqnarray}
\hspace{9mm} and
\begin{eqnarray}
\sum_{k=0}^{\infty} |( x|\psi_{n})|^{2} &\leq& r_{\psi} \| x \|^{2}, \nonumber
\end{eqnarray}
\hspace{9mm} for all elements $x$ of ${\cal H}$.\\
\hspace{3mm} (iii) $S_{\phi}$ and $S_{\psi}$ are bounded operators on ${\cal H}$.}\\
\par
{\it Proof.} (i)$\Rightarrow$(iii) This follows from Lemma 2.9, (3).\\
(iii)$\Rightarrow$(ii) Take an arbitrary $x \in {\cal H}$. Then we have
\begin{eqnarray}
\sum_{n=0}^{\infty} |(x| \phi_{n})|^{2}
&=& (S_{\phi} x |x) \nonumber\\
&\leq& \| S_{\phi} \| \|x \|^{2} , \nonumber
\end{eqnarray}
and
\begin{eqnarray}
\sum_{n=0}^{\infty} |(x| \psi_{n})|^{2}
&=& (S_{\psi} x |x) \nonumber\\
&\leq& \| S_{\psi} \| \|x \|^{2}. \nonumber
\end{eqnarray}
Hence, $\{ \phi_{n} \}$ and $\{ \psi_{n} \}$ are Bessel sequences.\\
(ii)$\Rightarrow$(i) Take an arbitrary $x = \sum_{k=0}^{n} (x|e_{k})e_{k} \in D_{\bm{e}}$. Then we have
\begin{eqnarray}
|(T_{\bm{e}}x|y)|
&=&
\left| \sum_{k=0}^{n} (x|e_{k})(\phi_{k}|y) \right| \nonumber \\
&\leq& \left( \sum_{k=0}^{n} |(x|e_{k})|^{2} \right)^{\frac{1}{2}} \left( \sum_{k=0}^{n} |(\phi_{k} |y)|^{2} \right)^{\frac{1}{2}} \nonumber \\
&\leq& \| x\| \left( \sum_{k=0}^{\infty} |(\phi_{k} |y)|^{2} \right)^{\frac{1}{2}} \nonumber \\
&\leq& \| x \| \left( r_{\phi}^{\frac{1}{2}} \| y \| \right) \nonumber
\end{eqnarray}
for each $y \in {\cal H}$, which implies that $T_{\bm{e}} \in B({\cal H})$. Similarly, we have $(T_{\bm{e}}^{-1})^{\ast} \in B({\cal H})$.
This completes the proof. \\\\
Furthermore, we have the following \\
\par
{\it Proposition 2.11.} {\it Let $( \{ \phi_{n} \}, \{ \psi_{n} \})$ be a regular biorthogonal pair. Then the following statements are equivalent:\\
\hspace{3mm} (i) $( \{ \phi_{n} \}, \{ \psi_{n} \})$ is a pair of semi-Riesz bases, that is, $T_{\bm{e}}$ (or $T_{\bm{e}}^{-1}$) is bounded.\\
\hspace{3mm} (ii) $\{ \phi_{n} \}$ (or $\{ \psi_{n} \}$) is a Bessel sequence. \\
\hspace{3mm} (iii) $S_{\phi}$ (or $S_{\psi}$) is a bounded operator on ${\cal H}$.}\\
\par
{\it Proof.} The statements (i), (ii) and (iii) are proved similarly to (i), (ii) and (iii) in Proposition 2.10, respectively.\\\\
The notion of Bessel sequences in Proposition 2.10 and 2.11 has appeared in Ref. \cite{bagarello10}.
\section{Physical operators defined by regular biorthogonal sequences}
In this section, let $( \{ \phi_{n} \}, \{ \psi_{n} \})$ be a regular biorthogonal pair and we study the following operators defined by $( \{ \phi_{n} \}, \{ \psi_{n} \})$
\begin{eqnarray}
A_{\bm{e}}
&=& T_{\bm{e}} \left( \sum_{k=0}^{\infty} \sqrt{k+1} e_{k} \otimes \bar{e}_{k+1} \right) T_{\bm{e}}^{-1} , \nonumber \\
B_{\bm{e}}
&=& T_{\bm{e}} \left( \sum_{k=0}^{\infty} \sqrt{k+1} e_{k+1} \otimes \bar{e}_{k} \right) T_{\bm{e}}^{-1}, \nonumber
\end{eqnarray}
where $T_{\bm{e}}$ is a closed operator for an ONB $\bm{e}= \{ e_{n} \}$ defined in Theorem 2.3. \\
\par
{\it Proposition 3.1.} {\it The following statements hold.\\
\hspace{3mm} (1) 
\begin{align}
A_{\bm{e}} \phi_{n} &= \left\{
\begin{array}{cl}
& 0, \;\;\; n =0  \\
& \sqrt{n} \phi_{n-1}, \;\;\; n\geq 1 . \\
\end{array}
\right. \nonumber 
\end{align}
\hspace{3mm} (2) $B_{\bm{e}} \phi_{n} = \sqrt{n+1} \phi_{n+1}$,   $n \geq 0$}\\
\par
{\it Proof.} (1) Since
\begin{align}
A_{\bm{e}} \phi_{n}
&=  T_{\bm{e}} \left( \sum_{k=0}^{\infty} \sqrt{k+1} e_{k} \otimes \bar{e}_{k+1} \right) T_{\bm{e}}^{-1} T_{\bm{e}} e_{n} \nonumber \\
&= T_{\bm{e}} \left( \sum_{k=0}^{\infty} \sqrt{k+1} (e_{n}|e_{k+1})e_{k} \right) , \nonumber 
\end{align}
we have
\begin{align}
A_{\bm{e}} \phi_{0}
&= 0, \nonumber \\
A_{\bm{e}} \phi_{n}
&= \sqrt{n} T_{\bm{e}} e_{n-1} = \sqrt{n} \phi_{n-1}, \;\;\; n=1,2, \cdots. \tag{3.1}
\end{align}
(2) This follows from
\begin{align}
B_{\bm{e}} \phi_{n}
&=  T_{\bm{e}} \left( \sum_{k=0}^{\infty} \sqrt{k+1} e_{k+1} \otimes \bar{e}_{k} \right) T_{\bm{e}}^{-1} T_{\bm{e}} e_{n} \nonumber \\
&= T_{\bm{e}} \left( \sum_{k=0}^{\infty} \sqrt{k+1} (e_{n}|e_{k})e_{k+1} \right) \nonumber \\
&= \sqrt{n+1} T_{\bm{e}}e_{n+1} \nonumber \\
&= \sqrt{n+1}\phi_{n+1}, \;\;\; n=0,1, \cdots . \tag{3.2}
\end{align}
These operators $A_{\bm{e}}$ and $B_{\bm{e}}$ are lowering and raising operators, respectively. These operators connect with ${\it quasi}$-${\it hermitian \; quantum \;  mechanics}$ and its relatives. We investigate the properties of $A_{\bm{e}}$ and $B_{\bm{e}}$, and the relationships between these operators and $T_{\bm{e}}$. In Theorem 2.3, we have shown that there uniquely exists an ONB $\bm{f}= \{ f_{n} \}$ in ${\cal H}$ such that $T_{\bm{f}}$ is a non-singular positive self-adjoint operator in ${\cal H}$.\\
\par
{\it Theorem 3.2.} {\it We have
\begin{eqnarray}
A_{\bm{e}}=A_{\bm{f}} \;\;\; {\rm and} \;\;\;
B_{\bm{e}}=B_{\bm{f}} \nonumber
\end{eqnarray}}
for any ONB $\bm{e}= \{ e_{n} \}$ in ${\cal H}$.\\
\par
{\it Proof.} Since $T_{\bm{e}}=T_{\bm{f}} U_{\bm{e},\bm{f}}$ by Lemma 2.4, we have
\begin{eqnarray}
A_{\bm{e}}
&=& T_{\bm{f}} U_{\bm{e},\bm{f}} \left( \sum_{k=0}^{\infty} \sqrt{k+1} e_{k} \otimes \bar{e}_{k+1} \right)U_{\bm{e}, \bm{f}}^{\ast} T_{\bm{f}}^{-1} \nonumber \\
&=& T_{\bm{f}} \left( \sum_{k=0}^{\infty} \sqrt{k+1} f_{k} \otimes \bar{f}_{k+1} \right) T_{\bm{f}}^{-1} \nonumber \\
&=& A_{\bm{f}}. \nonumber
\end{eqnarray}
It is similarly shown that $B_{\bm{e}}=B_{\bm{f}}$. This completes the proof.\\\\
This means that the lowering operators and the raising operators defined by ONB do not depend on methods of taking ONB. Therefore, we may consider only $A_{\bm{f}}$ and $B_{\bm{f}}$ as lowering and raising operators defined by ONB $\bm{f}= \{ f_{n} \}$ without loss of generality. We next define the lowering operator $B_{\bm{f}}^{\dagger}$ and the raising operator $A_{\bm{f}}^{\dagger}$ determined by $\{ \psi_{n} \}$ as follows:
\begin{eqnarray}
A_{\bm{f}}^{\dagger}
&=& T_{\bm{f}}^{-1} \left( \sum_{k=0}^{\infty} \sqrt{k+1} f_{k+1} \otimes \bar{f}_{k} \right) T_{\bm{f}}, \nonumber\\
B_{\bm{f}}^{\dagger}
&=& T_{\bm{f}}^{-1} \left( \sum_{k=0}^{\infty} \sqrt{k+1} f_{k} \otimes \bar{f}_{k+1} \right) T_{\bm{f}}.  \nonumber
\end{eqnarray}
Then we have the following \\
\par
{\it Proposition 3.3.} {\it The following statements hold.\\
\hspace{3mm} (1) 
\begin{eqnarray}
A_{\bm{f}}^{\dagger} \psi_{n}
&=& \sqrt{n+1} \psi_{n+1}, \;\;\;\;\;\;\;\; n=0,1, \cdots , \nonumber \\
B_{\bm{f}}^{\dagger} \psi_{n}
&=& \left\{
\begin{array}{cl}
& 0, \;\;\;\;\;\;\;\;\;\;\;\;\;\;\;\;\;  n=0 , \\
\nonumber \\
& \sqrt{n} \psi_{n-1}, \;\;\;\;\;\;\; n=1,2, \cdots . \\
\end{array}
\right. \nonumber 
\end{eqnarray}
\hspace{3mm} (2) $A_{\bm{f}}^{\dagger} \subset A_{\bm{f}}^{\ast}$ and $B_{\bm{f}}^{\dagger} \subset B_{\bm{f}}^{\ast}$.}\\
\par
{\it Proof.} (1) This is shown similarly to Proposition 3.1.\\
(2) Take an arbitrary $x \in D_{\bm{f}}$ such that $T_{\bm{f}} x \in D(A_{\bm{f}})$ and $y \in D( A_{\bm{f}}^{\dagger})$. Since 
\begin{eqnarray}
\left( \sum_{k=0}^{\infty} \sqrt{k+1} f_{k+1} \otimes \bar{f}_{k} \right)^{\ast}
= \sum_{k=0}^{\infty} \sqrt{k+1} f_{k} \otimes \bar{f}_{k+1} \nonumber
\end{eqnarray}
and $x \in D \left(T_{\bm{f}} \left( \sum_{k=0}^{\infty} \sqrt{k+1} f_{k} \otimes \bar{f}_{k+1} \right) \right)$, it follows that
\begin{eqnarray}
(T_{\bm{f}}x | A_{\bm{f}}^{\dagger} y )
&=& \left( x \; \left| \left( \sum_{k=0}^{\infty} \sqrt{k+1} f_{k+1} \otimes \bar{f}_{k} \right) \right. T_{\bm{f}} y \right) \nonumber \\
&=& \left( T_{\bm{f}} \left( \sum_{k=0}^{\infty} \sqrt{k+1} f_{k} \otimes \bar{f}_{k+1} \right) x \mid y \right)  \nonumber \\
&=& (A_{\bm{f}}T_{\bm{f}}x \; | y) , \nonumber
\end{eqnarray}
which implies that $A_{\bm{f}}^{\dagger} \subset A_{\bm{f}}^{\ast}$. Similarly we have $B_{\bm{f}}^{\dagger} \subset B_{\bm{f}}^{\ast}$. This completes the proof.\\\\
For relations between a regular biorthogonal pair $( \{ \phi_{n} \}, \{ \psi_{n} \})$ and lowering operators $A_{\bm{f}}, \; B_{\bm{f}}^{\dagger}$ and raising operators $B_{\bm{f}}, \; A_{\bm{f}}^{\dagger}$, we have following
\\
\par
{\it Proposition 3.4.} {\it The following statements hold.\\
\hspace{3mm} (1) 
\begin{eqnarray}
\phi_{n}&=& \frac{1}{\sqrt{n !}} B_{\bm{f}}^{n} \phi_{0}, \;\;\; n=0,1, \cdots , \nonumber \\
\psi_{n}&=& \frac{1}{\sqrt{n!}} (A_{\bm{f}}^{\dagger})^{n} \psi_{0}, \;\;\; n=0,1, \cdots . \nonumber
\end{eqnarray}
\hspace{3mm} (2) 
\begin{eqnarray}
A_{\bm{f}} D_{\phi}
= D_{\phi}, \;\;\; B_{\bm{f}}D_{\phi} =D_{\phi} . \nonumber
\end{eqnarray}
\hspace{9mm} and
\begin{eqnarray}
A_{\bm{f}}^{\dagger} D_{\psi}
= D_{\psi}, \;\;\;
B_{\bm{f}}^{\dagger} D_{\psi}
=D_{\psi} . \nonumber
\end{eqnarray}}
\par
{\it Proof.} (1) This is easily shown by the definition of $A_{\bm{f}}^{\dagger}$ and $B_{\bm{f}}$.\\
(2) This follows from Proposition 3.1 and the above (1).\\\\
By Proposition 3.4, (2) we have $D_{\phi} \subset D(A_{\bm{f}}) \cap D(B_{\bm{f}})$ and $D_{\psi} \subset D(A_{\bm{f}}^{\dagger}) \cap D(B_{\bm{f}}^{\dagger})$. We consider when $D_{\phi}$ is a core for $\bar{A}_{\bm{f}}$ and $\bar{B}_{\bm{f}}$ and $D_{\psi}$ is a core for $\bar{A}_{\bm{f}}^{\dagger}$ and $\bar{B}_{\bm{f}}^{\dagger}$.
\\
\par
{\it Proposition 3.5.} {\it The following statements hold.\\
\hspace{3mm} (1) If $T_{\bm{f}}$ is bounded, then $D_{\phi}$ is a core for $\bar{A}_{\bm{f}}$ and $\bar{B}_{\bm{f}}$.\\
\hspace{3mm} (2) If $T_{\bm{f}}^{-1}$ is bounded, then $D_{\psi}$ is a core for $\bar{A}_{\bm{f}}^{\dagger}$ and $\bar{B}_{\bm{f}}^{\dagger}$.}\\
\par
{\it Proof.} We show that if $T_{\bm{f}}$ is bounded, $D_{\phi}$ is a core for $\bar{A}_{\bm{f}}$. Take an arbitrary $y \in D(A)$. Since $T_{\bm{f}}$ is bounded, there exists an element $x$ of ${\cal H}$ such that
\begin{eqnarray}
\lim_{n \rightarrow \infty} \left( \sum_{k=0}^{n} \sqrt{k+1} f_{k} \otimes \bar{f}_{k+1} \right) x
= \sum_{k=0}^{\infty} \sqrt{k+1} (x | f_{k+1})f_{k} \nonumber 
\end{eqnarray}
exists in ${\cal H}$ and $y=T_{\bm{f}}x$. We put
\begin{eqnarray}
x_{n} = \sum_{k=0}^{n} (x|f_{k}) f_{k} , \;\;\; n=0,1, \cdots. \nonumber
\end{eqnarray}
Then we have $\{ x_{n} \} \subset D_{\bm{f}}$, $ \lim_{n \rightarrow \infty} x_{n}=x$ and
\begin{align}
\lim_{n \rightarrow \infty} \left( \sum_{k=0}^{\infty} \sqrt{k+1} f_{k} \otimes \bar{f}_{k+1} \right) x_{n} 
&= \lim_{n \rightarrow \infty} \left( \sum_{k=0}^{\infty} \sqrt{k+1} f_{k} \otimes \bar{f}_{k+1} \right) \left( \sum_{j=0}^{n} (x|f_{j}) f_{j} \right) \nonumber \\
&= \lim_{n \rightarrow \infty} \sum_{k=0}^{\infty} \sum_{j=0}^{n} \sqrt{k+1} (x|f_{j})(f_{j}|f_{k+1})f_{k} \nonumber \\
&= \lim_{n \rightarrow \infty} \sum_{k=0}^{n-1} \sqrt{k+1}(x|f_{k+1})f_{k} \nonumber \\
&= \left( \sum_{k=0}^{\infty} \sqrt{k+1} f_{k} \otimes \bar{f}_{k+1} \right) x. \tag{3.3}
\end{align}
We here define a sequence $\{ y_{n} \}$ in $D_{\phi}$ by $y_{n}=T_{\bm{f}}x_{n}, \;\;\; n=0,1, \cdots$. Then since $T_{\bm{f}}$ is bounded, it follows that $\lim_{n \rightarrow \infty} y_{n}=T_{\bm{f}}x=y$ and by (3.3)
\begin{eqnarray}
\lim_{n \rightarrow \infty}A_{\bm{f}} y_{n} 
&=& \lim_{n \rightarrow \infty} T_{\bm{f}} \left( \sum_{k=0}^{\infty} \sqrt{k+1} f_{k} \otimes \bar{f}_{k+1} \right) T^{-1}_{\bm{f}} T_{\bm{f}} x_{n} \nonumber \\
&=& \lim_{n \rightarrow \infty} T_{\bm{f}} \left( \sum_{k=0}^{\infty} \sqrt{k+1} f_{k} \otimes \bar{f}_{k+1} \right) x_{n} \nonumber \\
&=& T_{\bm{f}} \left( \sum_{k=0}^{\infty} \sqrt{k+1} f_{k} \otimes \bar{f}_{k+1} \right) x \nonumber \\
&=& T_{\bm{f}} \left( \sum_{k=0}^{\infty} \sqrt{k+1} f_{k} \otimes \bar{f}_{k+1} \right) T^{-1}_{\bm{f}} T_{\bm{f}} x \nonumber \\
&=& A_{\bm{f}} y, \nonumber
\end{eqnarray}
which implies that $D_{\phi}$ is a core for $\bar{A}_{\bm{f}}$. The others are similarly shown. This completes the proof.\\\\
Next we consider the operators $A_{\bm{f}}B_{\bm{f}}$, $B_{\bm{f}}A_{\bm{f}}$, $A_{\bm{f}}^{\dagger} B_{\bm{f}}^{\dagger}$ and $B_{\bm{f}}^{\dagger}A_{\bm{f}}^{\dagger}$. We have the following \\
\par
{\it Lemma 3.6.} {\it The following statements hold.\\
\hspace{3mm} (1) 
\begin{eqnarray}
A_{\bm{f}}B_{\bm{f}} 
&\subset& T_{\bm{f}} \left( \sum_{k=0}^{\infty}  (k+1) f_{k} \otimes \bar{f}_{k} \right) T^{-1}_{\bm{f}}, \nonumber \\
B_{\bm{f}}A_{\bm{f}}
&\subset& T_{\bm{f}} \left( \sum_{k=0}^{\infty}  (k+1) f_{k+1} \otimes \bar{f}_{k+1} \right) T^{-1}_{\bm{f}}. \nonumber
\end{eqnarray}
\hspace{3mm} (2) 
\begin{eqnarray}
B^{\dagger}_{\bm{f}}A^{\dagger}_{\bm{f}}
&\subset& T_{\bm{f}}^{-1} \left( \sum_{k=0}^{\infty}  (k+1) f_{k} \otimes \bar{f}_{k} \right) T_{\bm{f}}, \nonumber \\
A^{\dagger}_{\bm{f}}B^{\dagger}_{\bm{f}} 
&\subset& T_{\bm{f}}^{-1} \left( \sum_{k=0}^{\infty}  (k+1) f_{k+1} \otimes \bar{f}_{k+1} \right) T_{\bm{f}}. \nonumber 
\end{eqnarray}}
\par
{\it Proof.} (1) This follows from
\begin{eqnarray}
D(A_{\bm{f}} B_{\bm{f}})
&=& \left\{ x \in D(T_{\bm{f}}^{-1}); T_{\bm{f}}^{-1} x \in D \left(  \left( \sum_{k=0}^{\infty} \sqrt{k+1} f_{k+1} \otimes \bar{f}_{k} \right) \right)  \right. \nonumber \\ 
&&\hspace*{20mm} {\rm and} \;  \left( \sum_{k=0}^{\infty} \sqrt{k+1} f_{k+1} \otimes \bar{f}_{k} \right) T_{\bm{f}}^{-1} x \in D(T_{\bm{f}}) \nonumber \\
&&\hspace*{20mm}\left. {\rm and} \;  \left( \sum_{k=0}^{\infty}  (k+1) f_{k} \otimes \bar{f}_{k} \right) T_{\bm{f}}^{-1} x \in D(T_{\bm{f}}) \right\}  \nonumber \\
\nonumber \\
&\subset& \left\{ x \in D(T_{\bm{f}}^{-1}); \left( \sum_{k=0}^{\infty}  (k+1) f_{k} \otimes \bar{f}_{k} \right) T_{\bm{f}}^{-1} x \in D(T_{\bm{f}}) \right\} \nonumber \\
&=& D \left( T_{\bm{f}} \left( \sum_{k=0}^{\infty}  (k+1) f_{k} \otimes \bar{f}_{k} \right) T^{-1}_{\bm{f}} \right), \nonumber
\end{eqnarray}
and
\begin{eqnarray}
D(B_{\bm{f}} A_{\bm{f}})
&=& \left\{ x \in D(T_{\bm{f}}^{-1}); T_{\bm{f}}^{-1} x \in D \left(  \left( \sum_{k=0}^{\infty} \sqrt{k+1} f_{k} \otimes \bar{f}_{k+1} \right) \right)  \right. \nonumber \\ 
&&\hspace*{20mm} {\rm and} \;  \left( \sum_{k=0}^{\infty} \sqrt{k+1} f_{k} \otimes \bar{f}_{k+1} \right) T_{\bm{f}}^{-1} x \in D(T_{\bm{f}}) \nonumber \\
&&\hspace*{20mm}\left. {\rm and} \;  \left( \sum_{k=0}^{\infty}  (k+1) f_{k+1} \otimes \bar{f}_{k+1} \right) T_{\bm{f}}^{-1} x \in D(T_{\bm{f}}) \right\}  \nonumber \\
\nonumber \\
&\subset& \left\{ x \in D(T_{\bm{f}}^{-1}); \left( \sum_{k=0}^{\infty}  (k+1) f_{k+1} \otimes \bar{f}_{k+1} \right) T_{\bm{f}}^{-1} x \in D(T_{\bm{f}}) \right\} \nonumber \\
\nonumber \\
&=& D \left( T_{\bm{f}} \left( \sum_{k=0}^{\infty}  (k+1) f_{k+1} \otimes \bar{f}_{k+1} \right) T^{-1}_{\bm{f}} \right) . \nonumber 
\end{eqnarray}
(2) This is shown similarly to (1).\\\\
By Lemma3.6, we have the following\\
\par
{\it Theorem 3.7.} {\it The following statements hold.\\
\hspace{3mm} 
\begin{eqnarray}
A_{\bm{f}}B_{\bm{f}} - B_{\bm{f}}A_{\bm{f}} \subset I \;\;\;{\rm  and} \;\;\;
B_{\bm{f}}^{\dagger} A_{\bm{f}}^{\dagger} -A_{\bm{f}}^{\dagger}B_{\bm{f}}^{\dagger} \subset I . \nonumber
\end{eqnarray}}
\\
We finally define number operators $N_{\bm{f}}$ and $N_{\bm{f}}^{\dagger}$ by
\begin{eqnarray}
N_{\bm{f}}
&=& T_{\bm{f}} \left( \sum_{k=0}^{\infty} \sqrt{k+1} f_{k+1} \otimes \bar{f}_{k+1} \right) T_{\bm{f}}^{-1}, \nonumber \\
N_{\bm{f}}^{\dagger}
&=& T_{\bm{f}}^{-1} \left( \sum_{k=0}^{\infty} \sqrt{k+1} f_{k+1} \otimes \bar{f}_{k+1} \right) T_{\bm{f}}. \nonumber 
\end{eqnarray}
By Lemma 3.6, we have following\\
\par
{\it Proposition 3.8.} {\it The following statements hold.\\
\hspace{3mm} (1) 
\begin{eqnarray}
N_{\bm{f}} \phi_{n}
&=&  n \phi_{n}, \;\;\; n=0,1, \cdots, \nonumber \\
N_{\bm{f}}^{\dagger} \psi_{n}
&=& n \psi_{n}, \;\;\; n=0,1, \cdots . \nonumber
\end{eqnarray}
\hspace{3mm} (2) 
\begin{eqnarray}
 B_{\bm{f}}A_{\bm{f}} \subset N_{\bm{f}} \;\;\; {\rm and} \;\;\;  N_{\bm{f}}^{\dagger} \subset A_{\bm{f}}^{\dagger}B_{\bm{f}}^{\dagger}  . \nonumber
\end{eqnarray}}
\par
{\it Proof.} This follows from Lemma 3.6.

\ \\
Graduate School of Mathematics, Kyushu University, 744 Motooka, Nishi-ku, Fukuoka 819-0395, Japan
\\
h-inoue@math.kyushu-u.ac.jp,

\end{document}